\documentstyle[epsfig,cite,rotating,12pt]{article}

\begin{document}



\newcommand{\bfx}{ {\bf x} }
\newcommand{\cov}{ {\rm cov}}
\newcommand{\var}{ {\rm var}}

\newcommand{\rd}{{\mbox{d}}}
\newcommand{\rdd}{{\mbox{{\footnotesize d}}}}
\newcommand{\rds}{{\mbox{{\small d}}}}
\newcommand{\dx}{{\mbox{d}}x}
\newcommand{\dy}{{\mbox{d}}y}
\newcommand{\dz}{{\mbox{d}}z}
\newcommand{\lt}{<}
\newcommand{\gt}{>}
\newcommand{\beq}{\begin{equation}}
\newcommand{\eeq}{\end{equation}  }
\def\simgr{^>\hskip -2.5mm_\sim}
\def\simkl{^<\hskip -2.5mm_\sim}
\newcommand{\bec}{\begin{center}}
\newcommand{\eec}{\end{center}}
\def\b{\beta}
\def\g{\gamma}
\def\d{\delta}
\def\D{\Delta}
\def\e{\epsilon}
\def\as{\alpha_s}
\def\n{\nu}
\def\va{\varphi}
\def\p{\pi}
\def\r{\rho}
\def\E{\mbox{e}^+\mbox{e}^-}
\def\ran{\rangle}
\def\lan{\langle}

\def\Z{\mbox{Z}^o}
\def\qq{\mbox{q}\bar{\mbox{q}}}
\def\S{\sqrt{s}}

\def\Th{\Theta}
\def\Th0{\Theta_0}

\vspace{3.5cm}

\begin{center}

{\bf
ANGULAR INTERMITTENCY AND ANALYTICAL QCD PREDICTIONS
}

\vspace{1.0cm}

S.V.Chekanov\footnote[1]{On leave from
Institute of Physics,  AS of Belarus,
Skaryna av.70, Minsk 220072, Belarus.}

{\it High Energy Physics Institute Nijmegen
(HEFIN),  University  of Nijmegen/NIKHEF, \\
NL-6525 ED Nijmegen,  The Netherlands}

\vspace{1.0cm}

{\bf For  the L3 Collaboration}

\vspace{1.0cm}

Presented at 32th Rencontres de Moriond, \\
``QCD and High Energy Hadronic Interactions'' \\
Les Arcs, France, 1997


\vspace{1.0cm}

\begin{abstract}
We present a
comparison of local  multiplicity  fluctuations
in angular phase-space intervals  
with  first-order QCD  predictions.  
The data are based on  810k hadronic events
at $\S\simeq 91.2$~GeV  collected with  the L3 detector at LEP
during 1994.  
\end{abstract}

\end{center}

\vspace{2.0cm}

Recently, progress has  been made to derive  analytical QCD predictions
for angular intermittency  \cite{ochs,drem,brax}. 
Attempts have  been undertaken by the 
DELPHI Collaboration 
to compare the  predictions of \cite{ochs} 
with the data for  hadronic $\Z$ decay \cite{del}.

In this paper we
extend this study and present a first quantitative  
comparison of the theoretical first-order QCD predictions
\cite{drem,brax} with the L3 data,
emphasizing  the behavior of  
normalized factorial moments of orders $q=2,\ldots,5$ 
in angular phase-space intervals.

QCD predictions have been obtained \cite{drem,brax} 
for normalized factorial
moments  (NFMs) $F_q(\Theta)=\langle n(n-1)\ldots
(n-q+1)\rangle /\langle n\rangle ^q$, which show following scaling behavior 
$F_q(\Theta)\propto(\Theta_0/\Theta )^{(D-D_q)(q-1)}$.
For the one-dimensional case  ($D=1$), 
$\Theta_0$ is the opening half angle of a cone  around the jet-axis,
$\Theta$ is the angular half-width window of 
rings around the jet-axis centered at $\Theta_0$ and $n$ is
the number of particles in this window.
QCD expectations for  $D_q$
are as follows \cite{drem,brax}:

\vspace{0.4cm}

1) In a fixed coupling regime of the  Double Leading Log Approximation (DLLA), 
$D_q=\gamma_0(Q)(q+1)/q$,
where $\gamma_0(Q)=\sqrt{2\,C_{\mathrm{A}}\as (Q)/\pi}$ is
the anomalous QCD dimension calculated at  $Q\simeq  E\Theta_0$,
$E=\sqrt{s}/2$, and 
$C_{\mathrm{A}}=3$ is the gluon color factor.
$\as (Q)$ is evaluated according to the first order QCD with $n_f=5$
flavors.

\vspace{0.4cm}

2) In a  running-coupling regime of DLLA,
the $D_q$ have the form
$$
D_q\simeq\gamma_0(Q)\frac{q+1}{q}\left(1 + \frac{q^2+1}{4q^2}z\right)
\quad (1), \quad
D_q\simeq2\,\gamma_0(Q)\frac{q+1}{q}\left(\frac{1-\sqrt{1-z}}{z}\right)
\quad (2),
$$
were $z=\ln (\Theta_0/\Theta )/\ln (E\Theta_0/\Lambda )$.
Expressions (1) and (2) were  obtained in \cite{drem} and \cite{brax},
respectively. 

\vspace{0.4cm}

3) In the Modified Leading Log Approximation (MLLA),  
(1)  remains valid, except that $\gamma_0(Q)$ is replaced by  an 
effective $\gamma_0^{\mathrm{eff}}(Q)$ 
depending on $q$ \cite{drem}. 

\vspace{0.6cm}

For our comparison of the data with the 
theoretical predictions quoted above,
we will use the following  parameters:
$\Theta_0=25^0,  \Lambda=0.16$~GeV .
The first parameter is free.
Its  value  is  chosen to make our study comparable with  the
DELPHI analysis \cite{del}.
The value of $\Lambda$ chosen is that found in
our most recent determination of
$\as (m_{\mathrm{Z}})$ \cite{alex}.

\begin{figure}[htbp]

\vspace{-1.5cm}
\begin{center}
\mbox{\epsfig{file=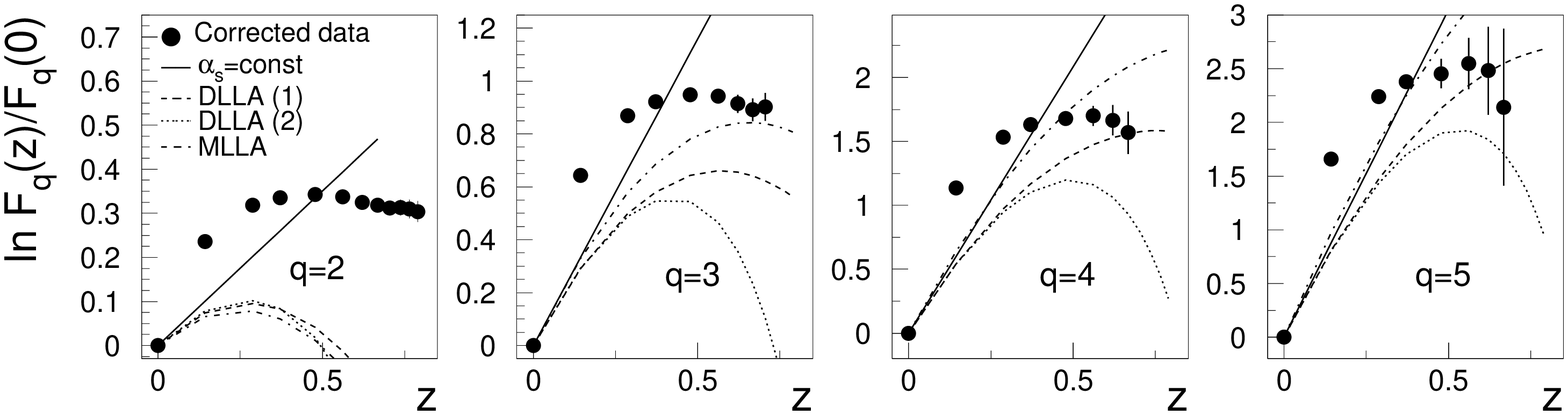, width=1.0\linewidth}}
\end{center}
\caption{
{\it The analytical  QCD predictions for $\Lambda=0.16\>\mathrm{GeV}$:
1) $\as=const$ ;
2) DLLA (eq. (1)); 3) DLLA (eq. (2));
4) MLLA.}}
\label{fa4}
\end{figure}

The comparison of the 
analytical QCD predictions 
to the  data  is  shown in Fig.~\ref{fa4}.
The data are corrected for detector imperfections, 
initial-state photon radiation,
Bose-Einstein correlations and Dalitz decays using Monte Carlo.
The predictions 
lead to the saturation effects seen 
in the data, but significantly underestimate 
the observed signal for $q=2$.
The reason for the saturation effect seen on the QCD predictions is
the dependence of $\as(Q)$
on $\Theta$. 
The fixed coupling regime  
(solid lines in  Fig.~\ref{fa4})  approximates 
the running coupling 
regime for small $z$, but does not exhibit 
the saturation effect seen in the data.
The MLLA predictions  do not  
differ significantly 
from the two DLLA  result for running coupling regime.

We have varied $\Lambda$ in the range of $0.04-0.25$~GeV. 
We found that the disagreement observed
is valid for relatively large values 
of $\Lambda$  as well
as for small values (down to $\Lambda =0.04$~GeV). In the latter case, 
a reasonable estimate for the
second-order NFM  can be reached, consistent with 
the DELPHI conclusion \cite{del}.
However, our analysis shows that, in this case, the theoretical
higher-order NFMs overestimate the data.

Note that the disagreement for the second-order NFMs
can  be reduced by considering the second-order expression for $\as (Q)$
or by replacing $n_f=3$,  instead of $n_f=5$.  This leads to a decrease of
the $\gamma_0(E\Theta_0)$. 
However, also in this case 
good agreement cannot  be achieved  for  higher-order   NFMs.

{\bf Conclusion.}
The analytical first-order  perturbative QCD predictions
are shown to be in disagreement with the local 
fluctuations  observed for hadronic $\Z$ decay.

{\bf Acknowledgments.}
We wish to express our gratitude to the CERN accelerator divisions for
the excellent performance of the LEP machine.  We acknowledge the
effort of all engineers and technicians who have participated in the
construction and maintenance of this experiment.
We thank  B.Buschbeck,  P.Lipa, 
V.I.Kuvshinov, J.-L.Meunier,  R.Peschanski and W.Ochs 
for useful discussions.
 
%

\end{document}